\documentclass[twocolumn,showpacs,aps,prd,nobibnotes,floatfix]{revtex4-1}
\usepackage{amsmath}
\usepackage[hmargin=1.5cm,vmargin=1.5cm,bindingoffset=0.0cm]{geometry}

\usepackage{graphicx}
\usepackage{longtable}
\usepackage{graphics,placeins,subfigure,wrapfig}
\usepackage[usenames]{color}
\usepackage{amsmath}
\usepackage{soul}
\usepackage[export]{adjustbox}

\newcommand{\beq}{\begin{equation}}
\newcommand{\eeq}{\end{equation}}
\newcommand{\beqn}{\begin{eqnarray}}
\newcommand{\eeqn}{\end{eqnarray}}

\usepackage{xcolor}

\definecolor{forestgreen}{RGB}{35,142,35}

\begin{document}
\title{Induced Spins from Scattering Experiments of Initially Nonspinning Black Holes}
\author{Patrick E. Nelson$^{1}$, Zachariah B. Etienne$^{1}$, Sean T. McWilliams$^{1}$, Viviana Nguyen$^{2}$}
\affiliation{$^1$Department of Physics and Astronomy, West Virginia
  University, Morgantown, West Virginia 26506, USA; Center for Gravitational Waves and Cosmology, West Virginia University, Morgantown, West Virginia 26506, USA}
\affiliation{$^2$ Department of Physics, University of Illinois at Urbana-Champaign, Urbana, Illinois 61801, USA}

\begin{abstract}

When two relativistically boosted, nonspinning black holes pass by one
another on a scattering trajectory, we might expect the tidal
interaction to spin up each black hole. We present the first exploration of
this effect, appearing at fourth post-Newtonian order, with
full numerical relativity calculations.
The basic setup 
for the calculations involves two free parameters: the initial boost of
each black hole and the initial angle between the velocity vectors and
a line connecting the centers of the black holes, with zero angle
corresponding to a head-on
trajectory. To minimize gauge effects, we measure final spins only if
the black holes reach a final separation of at least $20M$. Fixing the
initial boost, we find that as the initial angle decreases toward the
scattering/nonscattering limit, the
spin-up grows nonlinearly. In addition, as initial boosts are increased from
$0.42c$ to $0.78c$, the largest observed final dimensionless spin on
each black hole increases nonlinearly from $0.02$ to
$0.20$. Based on these results, we conclude that much higher
spin-ups may be possible with larger boosts, although achieving this will
require improved numerical techniques.
\end{abstract}

\maketitle

\section{Introduction}

Inelastic scattering experiments in high-energy
physics have deepened our understanding of nonlinear
interactions between quarks within hadrons and
mesons~\cite{Bloom:1969,Breidenbach:1969}. Analogously, we
would expect black hole scattering experiments performed in full
numerical relativity to provide insights
into strong-field gravity at its strongest and most dynamical. This
paper presents such experiments set up in a way that
effectively eliminates gauge ambiguities.
Specifically, we will address to what degree two
relativistically boosted, equal-mass, nonspinning black holes
on a scattering trajectory induce spins on each other.

The notion that a black hole's spin may be influenced by a
distant object is not a new one; for example in 1974, Hartle
\cite{Hartle:1974} demonstrated with analytical arguments that a
stationary, coplanar moon far from a spinning black hole will act to
spin down the hole. More relevant to our work, a similar effect was
studied by Campanelli, Lousto, and Zlochower for black hole binaries
in quasicircular orbits \cite{Campanelli:2006}. They also derived
a formula for the radiated angular momentum based on the Weyl scalar
$\psi_4$ \cite{Lousto:2007}. However, our work focuses on scattering systems.

Our experimental technique is as follows. We first uniquely specify
initial black hole trajectories with Brandt-Br{\"u}gmann initial data
parameters, then perform scattering experiments in full numerical
relativity, and finally measure the final
black hole spins at large separations. Final spins are measured with
both the isolated horizon
formalism~\cite{Ashtekar:2004L} and the Christodoulou
formula (involving the ratio of proper equatorial to polar
circumferences of the apparent horizons)~\cite{Alcubierre:2004}.

The final spins encode important information about the
strong-field interaction, which first appears
at fourth order in a post-Newtonian expansion in terms that account for the flux of
angular momentum at the event horizon
\cite{Poisson:1994}. This
work therefore describes a potential way
to validate post-Newtonian
calculations at very high order with numerical relativity.

Although this work appears to be the first to focus on the final spin of
scattering black hole encounters, it is not the first to
explore black holes on highly eccentric or hyperbolic trajectories. In one
class of such interactions, orbits undergo ``zooms'' and ``whirls''---that is, they whirl close to each other before zooming out to a larger
separation, in an extreme example of the precession of the peribothron
that occurs near the separatrix of bound and unbound orbits. This class of orbits has
been studied in stationary black hole
spacetimes~\cite{Chandrasekhar:579245,Martel:2003,Levin:2008}, in
the post-Newtonian limit~\cite{Grossman:2009,Levin:2009}, and in
extreme-mass-ratio systems ~\cite{Glampedakis:2002,Drasco:2005,Haas:2007,Hughes:2005}, and
is related to the homoclinic
orbits observed in stationary black hole spacetimes~\cite{Bombelli:1992} and in the post-Newtonian limit~\cite{Cornish:2003}. 
Khurana and Pretorius were
the first to show evidence of zoom-whirl orbits in full numerical relativity, for equal-mass,
nonspinning binaries. They simulated up to five zoom-whirls, but showed
that many more may be
possible, since the number of orbits exhibits critical
phenomenon behavior, with exponential scaling in the number of orbits $n$
with the critical impact parameter $b^*$:
$e^n \propto |b-b^*|^\gamma$~\cite{Pretorius:2007}. Although Khurana and 
Pretorius fine-tuned their initial data in an attempt to maximize the
number of zoom-whirls, Healy, Levin, and Shoemaker found up to three
orbits without fine-tuning~\cite{Healy:2009}. 

Other studies of zoom-whirl behavior in equal-mass, nonspinning
binaries have been performed (see, e.g., Refs.~\cite{Gold:2013, Gold:2010}.
These trajectories have also been
studied for comparable-mass Kerr black holes~\cite{Grossman:2009} and
large-mass-ratio Kerr black holes~\cite{Levin:2011}, as well as for
binary systems that contain one spinning black hole and one
nonspinning black hole~\cite{Levin:2009}. Zooms and whirls have even
been observed for  hyperbolic orbits. Gold and Br{\"u}gmann showed
orbits that exhibited a single whirl followed by a zoom to infinity
\cite{Gold:2012}. Our focus will be on highly eccentric encounters
similar to these, with the black holes reaching large separations
quickly after a close encounter. Our spin measures assume isolated
black holes, so to minimize potential gauge or strong-field effects on
these measurements, we only present final spin measures if, after a
strong-field encounter, the final separation of the black holes
remains at least $20M$ through the end of the calculation; we refer to
encounters that satisfy this requirement as 
``scattering'' events, whereas all other encounters are referred to as ``nonscattering'' or ``merger''
events. Throughout this work, we will work in geometrized units where $G=c=1$.

\begin{figure}[ht!]
\centering
\includegraphics[width=0.49\textwidth]{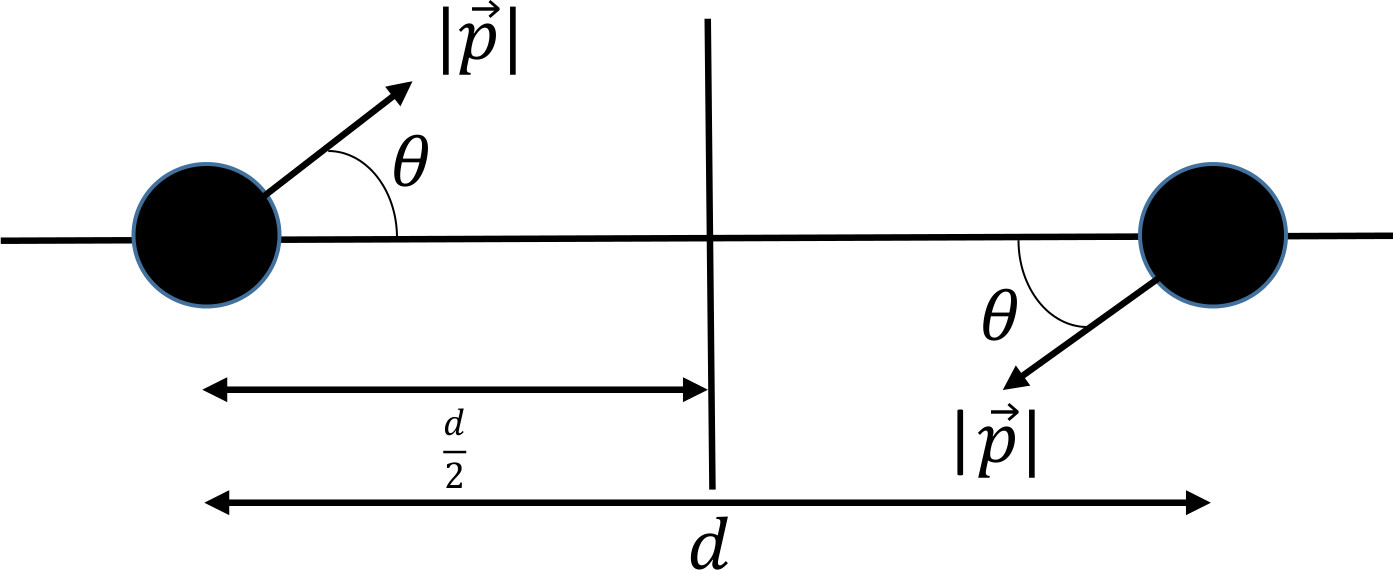}
\caption{Schematic of the chosen initial black hole
  configuration. $|\vec{p}|$ is the magnitude of the
  Brandt-Br{\"u}gmann momentum, $\theta$ is the shooting angle formed
  with the $x$ axis, and $d$ is the initial orbital separation. For
  calculations presented here, we choose $d=100$ in code units,
  corresponding to initial separations between $52.4M$ in the case of
  an initial $v=0.77$ boost and $96.6M$ in the case of an initial
  $v=0.42$ boost ($c=1$ units; $M$ is the initial ADM mass), as summarized in
  Table~\ref{tab:initialconditions}.}
\label{initialdataschematic}
\end{figure}

As illustrated in Fig.~\ref{initialdataschematic}, our experimental
setup is quite similar to those adopted in the zoom-whirl literature
for equal-mass black holes \cite{Pretorius:2007, Healy:2009,
  Gold:2013, Gold:2010}, with the key exception that we start with a
much larger separation of $100$ in code units. This translates to
between 50$M$ separation with the highest boosts chosen up to 96$M$
separation with the smallest initial boost, as the initial
Arnowitt-Deser-Misner (ADM) mass increases
monotonically with the initial boost. The actual values are summarized in
Table~\ref{tab:initialconditions}.  We find the choice of large
initial separations to be especially important, as it gives the gauge
fields ample time to settle and 
enables junk radiation to propagate away so that the 
gravitational-wave signal due to the dynamical interaction can be analyzed.

At fixed Brandt-Br{\"u}gmann
momentum magnitude $|\vec{p}|$, we find the induced spin-up increases
monotonically as $\theta$ decreases toward the
scattering/nonscattering separatrix.
In addition, the maximum measured 
spin-up is found to be 0.02, 0.06, 0.11, and 0.20 with initial
boosts of $v = 0.42$, 0.56, 0.66, and 0.78, respectively. This represents a
significantly nonlinear trend as the initial boost parameter is
increased. While the induced spins may not seem particularly large for
astrophysically motivated boosts such as those potentially seen in
hierarchical triple systems, we note that for bound systems, even
small induced spins in each encounter may accumulate and
significantly impact the dynamics over many orbital time scales.

The rest of the paper is organized as follows:
Sec.~\ref{methods} discusses the methods used for both the
simulations and data analysis; Sec.~\ref{ics} describes the initial
conditions of each simulation; Sec.~\ref{results} presents our results; and
Sec.~\ref{conc} contains our conclusions and plans for future work.

\section{Numerical Approach and Diagnostics}
\label{methods}

We adopt standard moving puncture techniques, combining the 3+1
Baumgarte-Shapiro-Shibata-Nakamura (BSSN) formalism
\cite{Shibata:1995, Baumgarte:1999}
with moving puncture gauge conditions to construct the spacetime. We
choose $W = e^{-2\phi}$ as our evolved conformal variable, and
standard {\tt 1+log} lapse and Gamma-driver shift conditions.

Initial data are set up and evolutions performed using open-source
tools within the Einstein Toolkit (ETK) \cite{Loffler:2012,
  EinsteinToolkitWebsite} infrastructure. In particular,
constraint-satisfying Brandt-Br{\"u}gmann two-black-hole initial data
are generated using the {\tt TwoPunctures} module~\cite{Ansorg:2004}
with optimized spectral interpolation
\cite{Paschalidis:2013}. The evolution of the BSSN equations was
performed with the {\tt McLachlan}
\cite{Brown:2008,Kranc:web,McLachlan:web} module. {\tt 
  QuasiLocalMeasures} was applied to calculate black hole spins
\cite{Dreyer:2003}, {\tt AHFinderDirect} to measure horizon centroids
and circumferences \cite{Thornburg:2004,Thornburg:1995}, {\tt
  WeylScal4} to compute $\Psi_4$ \cite{Zilhao:2013}, as well as the
Cactus Computational Toolkit
\cite{Goodale2003,CactusWebsite,Cactusprize:web} and Carpet
\cite{Schnetter:2004, CarpetWebsite} for the adaptive mesh refinement grid
infrastructure. 

\subsection{AMR Grid Structure}
\label{gridstructure}

The adaptive-mesh-refinement (AMR) grids generated by Carpet use half-side lengths of $0.75
\times 2^n$, for $n=\lbrace 0,\ldots,6 \rbrace,\lbrace 8,\ldots,10
\rbrace$ in code units. The skip between $n=6$ and $n=8$ ensures a
large region within which gravitational waves can be extracted at
uniform resolution. All
simulations enforce reflection symmetry across the orbital plane to
minimize computational expense. The outermost boundary has a
half-side length of 768, which ensures that approximate outer boundary
conditions can have no causal impact on the spin measures. The three
resolutions used for the most refined grid are $\Delta x = \lbrace
1/56,1/66.\overline{6},1/85.\overline{3} \rbrace;$ henceforth, these
shall be referred to as low, medium, and high resolution,
respectively. Note that these resolutions are also given in code
units, so that higher-boost cases will have higher resolutions
(the initial ADM mass $M$ for each boost is given in code units in Table~\ref{spincutoff}). For
boosts other than $0.66$, only the ``low'' resolution was
used. Finally, the Carpet parameter {\tt time\_refinement\_factors}, which
controls how often a time step is performed on each refinement level, is
set to {\tt  [1,1,1,1,2,4,8,16,32,64]}, so that the coarsest four grids are
updated at the same rate, and each refinement level finer than that is
updated twice as often as the next-coarser grid. This minimizes
interpolation errors on the coarsest levels by disabling time prolongation
between them.

\subsection{Dimensionless Spin}

The dimensionless spin of each black hole, $a/M$, is calculated using
two approaches. First is the Christodolou spin, which is given by solving
Eq.~5.2 of Ref.~\cite{Alcubierre:2004}, 
\begin{equation}
C_r = \frac{1+\sqrt{1-(a/M)^2}}{\pi} E\left( -\frac{(a/M)^2}{\left(1+\sqrt{1-(a/M)^2}\right)^2} \right),
\end{equation}
where $C_r = C_p/C_e$ is the ratio of the polar and equatorial horizon
circumferences, and $E(x)$ is the complete elliptic integral of the
second kind, 
\begin{equation}
E(k) = \int_{0}^{\pi/2} \sqrt{1-k\sin^2\theta}d\theta.
\end{equation}
Second, the isolated horizon formulation is adopted as an alternative
spin measure, which is provided by the ETK {\tt QuasiLocalMeasures}
module.

We also employ standard numerical surface integrals for evaluating the
ADM mass $M_{\rm ADM}$ and angular momentum $J_{\rm ADM}$ in the Cartesian basis
to measure mass and angular momentum lost to gravitational waves and
spin-ups. Additionally, we define the ``final angular momentum''
and ``final spin'' to be a time average over the values of angular
momentum and spin \textit{after} the black holes have reached the cutoff
separation $r_{\rm cutoff}=20M$ as they move away from each other after
the encounter. 

\begin{table*}[t]
\centering
\begin{tabular}{ |c|c||c|c|c|c||c| }
\hline
$|\vec{p}|$ & $M_{\rm ADM}$ & SDA @ $\tilde{r}_{\rm cutoff} = 5.0$
& SDA @ $\tilde{r}_{\rm cutoff} = 10.0$
& SDA @ $\tilde{r}_{\rm cutoff} = 15.0$ & SDA @ $\tilde{r}_{\rm cutoff} = 20.0$ & Maximum $\tilde{r}_{\rm merge}$ \\
\hline
$0.490$ & 1.13560 & 3.6 (3.4) & 4.4 (3.7) & 4.6 (3.7) & --- (3.8) & 3.41 \\
$0.735$ & 1.28385 & 3.8 (3.8) & 4.5 (4.0) & 5.0 (4.1) & --- (4.2) & 3.95 \\
$0.980$ & 1.46473 & 3.7 (3.8) & 4.4 (4.1) & 4.9 (4.5) & --- (4.5) & 4.36 \\
$1.500$ & 1.90902 & 3.8 (4.0) & 4.8 (4.4) & 5.6 (4.7) & --- (4.7) & 3.75 \\
\hline
\end{tabular}
\caption{Validation of $r_{\rm cutoff}$ choice. Larger choices of
  $r_{\rm cutoff}$ ensure that the black holes are sufficiently far
  apart for an accurate final spin measurement and that they are
  unlikely to merge. The columns are described as follows. $|\vec{p}|$ is the
  Brandt-Br{\"u}gmann initial momentum's magnitude, and $M_{\rm ADM}$
  is the initial ADM mass in code units. The next four columns give
  the mean number of significant digits of agreement (SDA) between the
  Christodolou spin calculated with that cutoff radius and the
  Christodolou spin calculated with $r_{\rm cutoff} = 20.0M_{\rm ADM}$; the
  em dashes in the $\tilde{r}_{\rm cutoff} = 20.0$ column indicate
  that the SDA between a measurement and itself is not meaningful. In
  parentheses, we give the mean SDA between the Christodolou and
  isolated horizon formalism dimensionless spin measures at points
  closest to the listed cutoff
  separation. For brevity we define tilded $r$ quantities in this
  table to be normalized by $M_{\rm ADM}$; e.g.,
  $\tilde{r}_{\rm cutoff} = r_{\rm cutoff}/M_{\rm ADM}$. Finally,
  ``Maximum $\tilde{r}_{\rm merge}$'' is the maximum separation
  observed between the two holes for cases in which, after a
  strong-field encounter, they ultimately
  merged at the given boost.}
\label{spincutoff}
\end{table*}

Our measures of the final spin parameter assume that each black hole is
far from another strong-field source and generally on an unbound
trajectory. To confirm that our black hole spin 
measures are reliable, we only include cases in which the holes reach
a final separation of at least $r_{\rm cutoff}$ through the end of the
calculation. In Table~\ref{spincutoff}, we compare the agreement of
Christodolou spins calculated with different cutoff radii with the spin
calculated at our chosen $r_{\rm cutoff}=20M$; we see good agreement between
cutoffs of $15M$ and $20M$, showing that we have a stable spin measurement.
We also compare the Christodolou
spin and the isolated horizon formalism measures of spin with varying
choices of $r_{\rm cutoff}$, recording at each $r_{\rm cutoff}$ and
for each initial boost the significant digits of agreement between the
two spin measures (these are the values in parentheses in Table~\ref{spincutoff}).
The table confirms that at our fiducial
$r_{\rm cutoff}=20M$, the independent black hole spin measures agree
to better than four significant digits across most cases. As a point of
comparison, we find that over all chosen boost magnitudes and angles,
no black holes return to merge after reaching about $r=5M$
separation. 

\subsection{Parametrizing the Initial Boost}
\label{sec:initialboost}

The initial Brandt-Br{\"u}gmann boost parameter $\vec{p}_{\rm BB}/m_{\rm BB}$
provides an unambiguous measure of the initial boost of each black
hole. However, we would expect 
that the junk radiation associated with
the assumption of conformal flatness in Brandt-Br{\"u}gmann initial
data increases as this momentum increases, acting to reduce the
momentum of each black hole prior to the interaction
more and more as we increase the initial boost \cite{Ruchlin:2014}.

To parametrize our runs in a way that is insensitive to the presence
of junk radiation, we define the initial boost to be the
coordinate speed of a {\it single} black hole imparted with the same
initial Brandt-Br{\"u}gmann momentum, in the limit $t\to \infty$. In
particular, at each initial momentum chosen, we perform a dedicated
numerical relativity
calculation of a single black hole with this initial momentum,
traveling along the $x$ axis. We choose the AMR grids to be
identical to the low-resolution experiments, except instead of two AMR
grid hierarchies tracking one black hole each, we only need a single
hierarchy to track the single black hole.

Plotting the position of the black hole in this calculation versus
time, we find that it starts from near zero speed (due to the initial
shift $\beta^i$ being zero) and accelerates
towards some constant speed. However, attempts at a linear
fit to the late-time data revealed that the black hole coordinate
acceleration on the numerical grids was still nonzero at the end of
the calculation, when we were forced to terminate to ensure a valid
AMR hierarchy. Therefore, instead of fitting the speed at the end of the calculation, we
consider model functions that have asymptotes so the entire time
series can be used to estimate the speed that our initial conditions
represent.

\begin{figure}[t!]
\centering
\includegraphics[width=0.49\textwidth]{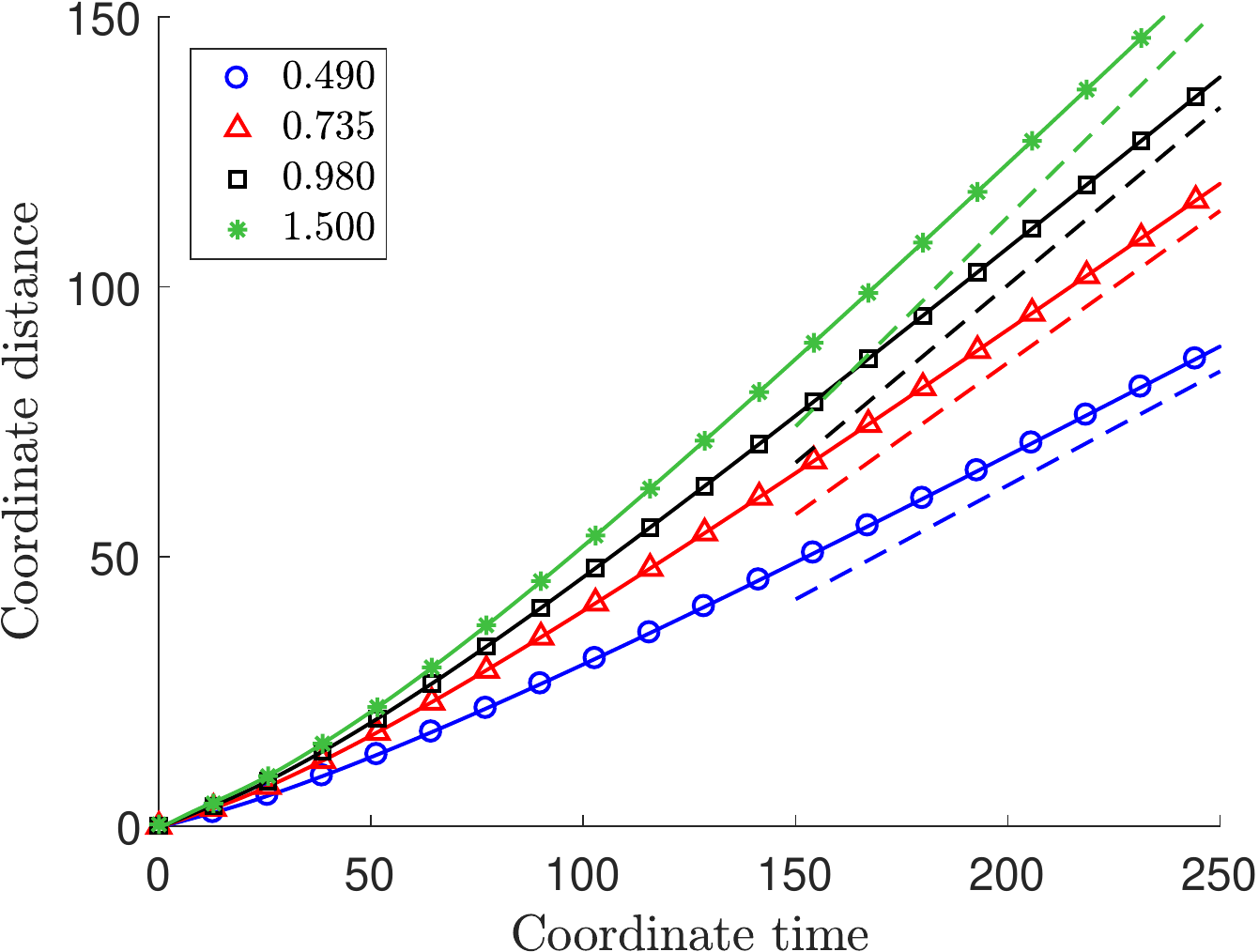}
\caption{Parametrizing the initial boost: Late-time data of dedicated
  single-black-hole runs to find the initial boost. Here, $t=0$
  corresponds to an arbitrarily chosen later time, and all of these
  boost calibration calculations were performed to a final time of
  $\approx 250M$. The dashed lines show the asymptotic slope, solid
  lines show the actual data from the numerical relativity calculation, and the points show the
  hyperbolic fit.}
\label{fig:initialboost}
\end{figure}

A hyperbola in the $x$-$t$ plane is a convenient choice here, so we
fit hyperbolae to the data by minimizing the residual norm cost
function
$$
f(a,b,h,k) = \sqrt{\sum_{i=1}^{N} \left( x_i - g(a,b,h,k,t_i) \right)^2},
$$ where $x_i$ and $t_i$ are the position and time of the
black hole at each of $N$ measurements of position, respectively.
$g(a,b,h,k,t)$ is the hyperbola
$$
g(a,b,h,k,t) = k - \sqrt{a^2 \left( 1 + \frac{(t-h)^2}{b^2} \right)},
$$
where $a$ is the semimajor axis, $b$ is the semiminor axis, and
$(h,k)$ are the coordinates of the center.
We then take the slope of the asymptote $a/b$ as the actual
preinteraction boost that corresponds to the input momentum. As
illustrated in Fig.~\ref{fig:initialboost}, the hyperbolic fit to data
plotted at these late times is quite good.

This hyperbolic fitting method produces
values of $v = 0.42$, 0.56, 0.66, and 0.78, corresponding to
magnitudes of the initial Brandt-Br{\"u}gmann momentum of $|\vec{p}| = 
0.49$, 0.735, 0.98, and 1.50, respectively. Interestingly, these estimates for
the initial boost are quite close to the respective values of $|\vec{p}|/M_{\rm ADM} =
0.43$, 0.57, 0.67, and 0.79. Notice however that our measured asymptotic
coordinate boost is always slightly lower than the Brandt-Br{\"u}gmann
momentum-based measure, presumably due to the emission of junk
radiation. 

\subsection{Radiated Angular Momentum}

As the black holes are initially nonspinning, the orbital angular
momentum contributes the entirety of the system's initial angular
momentum. As the numerical relativity calculation progresses, however, this
orbital contribution decreases as the black holes are spun up and
gravitational waves carry away angular momentum. The radiated angular
momentum $J_{\rm GW}$ is calculated using
\begin{equation}
J_{\rm GW}=\frac{r^2_{\rm ext}}{16 \pi}\sum_{l,m} \int_{-\infty}^t-m(\dot{h}_{+} h_{\times} - \dot{h}_{\times} h_{+}) dt ,
\label{eq:JGW}
\end{equation}
derived from Eq.~24 of Ref.~\cite{Campanelli:1999}. In all cases
we measure the strain from the outgoing Weyl scalar $\psi_4$ at
$r_{\rm ext} = 67.88M$, which is sufficiently far from the binary but
in a high-enough resolution region to yield a reliable result.

\section{Initial Conditions}
\label{ics}

For all sets of simulations, the initial separation was set to $d=100$.
Such large initial separations
allow more time for the junk-radiation-induced perturbation on
each black hole to settle before the holes strongly interact,
and ensure that a sufficient interval of time exists between the junk and the
gravitational radiation from the dynamical interaction. 

In a given simulation with $|\vec{p}|$ and $\theta$ (whether bound
or unbound), we set $p_x = \pm |\vec{p}| \cos (\theta)$ and $p_y =
\mp |\vec{p}| \sin (\theta)$, giving rise to the configuration
shown in Fig.~\ref{initialdataschematic} (this figure is not to
scale).

\begin{table*}[t]
\centering
\begin{tabular}{ |c|c|c|c|c|c|c|c| }
\hline
Boost ($v$) & $|\vec{p}|$ & Resolution & $d/M_{\rm ADM}$ & $\theta_{\rm N-S}$ & $N_{\theta_{\rm N-S}}$ & $\theta_{\rm S}$ & $N_{\theta_{\rm S}}$ \\
\hline
$0.42$ & $0.490$ & Low & 96.6 & $\{5.600 \times 10^{-2}, 6.040 \times 10^{-2}\}$ & $11$ & $\{6.050 \times 10^{-2}, 7.100 \times 10^{-2}\}$ & $23$ \\
$0.56$ & $0.735$ & Low & 77.9 & $\{5.400 \times 10^{-2}, 5.440 \times 10^{-2}\}$ & $3$ & $\{5.460 \times 10^{-2}, 6.500 \times 10^{-2}\}$ & $31$ \\
$0.66$ & $0.980$ & Low & 68.2 & $\{5.400 \times 10^{-2}, 5.420 \times 10^{-2}\}$ & $3$ & $\{5.430 \times 10^{-2}, 5.800 \times 10^{-2}\}$ & $28$ \\
$0.66$ & $0.980$ & Medium & 68.2 & $\{5.400 \times 10^{-2}, 5.420 \times 10^{-2}\}$ & $3$ & $\{5.430 \times 10^{-2}, 5.800 \times 10^{-2}\}$ & $28$ \\
$0.66$ & $0.980$ & High   & 68.2 & $\{5.418 \times 10^{-2}, 5.424 \times 10^{-2}\}$ & $3$ & $\{5.430 \times 10^{-2}, 5.690 \times 10^{-2}\}$ & $17$ \\
$0.77$ & $1.500$ & Low & 52.4 & $\{3.000 \times 10^{-2}, 5.950 \times 10^{-2}\}$ & $9$ & $\{5.960 \times 10^{-2}, 7.000 \times 10^{-2}\}$ & $21$ \\
\hline
\end{tabular}
\caption{Initial conditions for black hole scattering
  experiments. $|\vec{p}|$ is the magnitude of the initial
  Brandt-Br{\"u}gmann momentum; $\theta_{\rm N-S}$ and $\theta_{\rm S}$
  are the ranges of the initial shooting angles (in radians) of the
  Brandt-Br{\"u}gmann momentum for cases exhibiting scattering and
  nonscattering (``merging'') behavior, respectively; and
  $N_{\theta_{\rm N-S}}$ and $N_{\theta_{\rm S}}$  
  are the number of simulations in those ranges.}
\label{tab:initialconditions}
\end{table*}

Table~\ref{tab:initialconditions} presents a complete list of
calculations performed in this work. Notice that runs for
$|\vec{p}| = 0.980$ were carried out at three different
resolutions, with grids as specified in
Sec.~\ref{gridstructure}. Upon finding that results (i.e., final
measured spin parameters) were not significantly improved at higher
resolutions, other runs were carried out at low resolution.

The range of angles (given in radians) and number of runs
performed in each set of experiments are also given; these sets
are divided into subsets corresponding to bound and unbound trajectories. 
For instance, when setting the momentum magnitude to $0.490$, we find
that the initial boost is $v=0.42$ and that the angle
$\theta = 6.050 \times 10^{-2}$ marks the boundary between scattering
and nonscattering (i.e. ``merging'') cases. We performed 23 simulations
between that angle and $\theta = 7.100 \times 10^{-2}$ and an additional
11 simulations satisfying
$5.600 \times 10^{-2} \leq \theta \leq 6.040 \times 10^{-2}$ as we
searched for the transition between scattering 
and nonscattering trajectories. Note that the
separatrix angle between scattering/nonscattering is not monotonic in
boost; this happens because the fixed initial separation in code units
$d$ combined with the larger initial ADM mass $M$ in the higher-boost cases
decreases the initial dimensionless separation $d/M$, so the angles are not
immediately comparable.

\section{Results}
\label{results}

\subsection{Trajectory Morphologies}

\begin{figure*}[ht!]
\includegraphics[width=0.333\textwidth,valign=t]{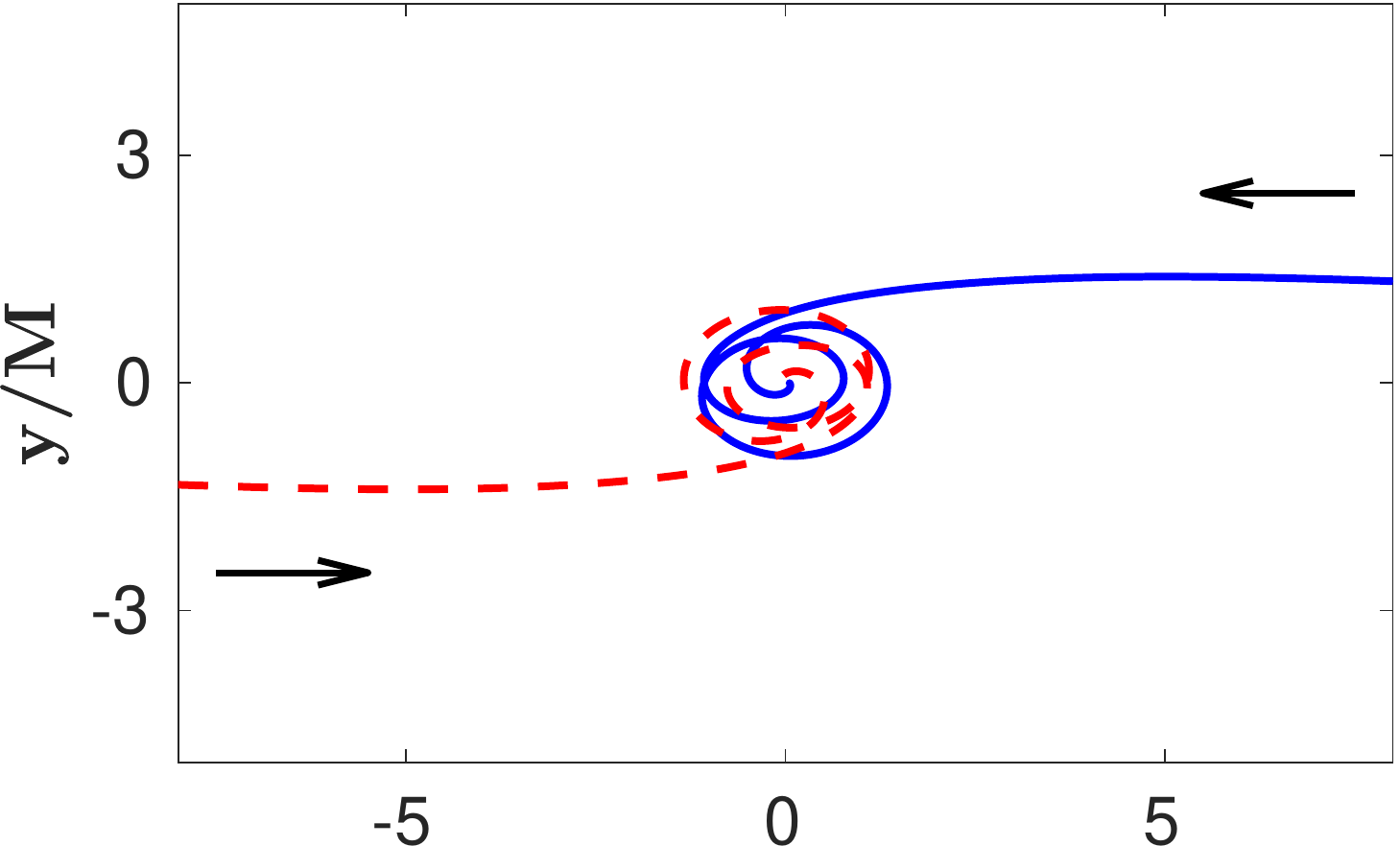}
\hspace{0 cm}
\includegraphics[width=0.299\textwidth,valign=t]{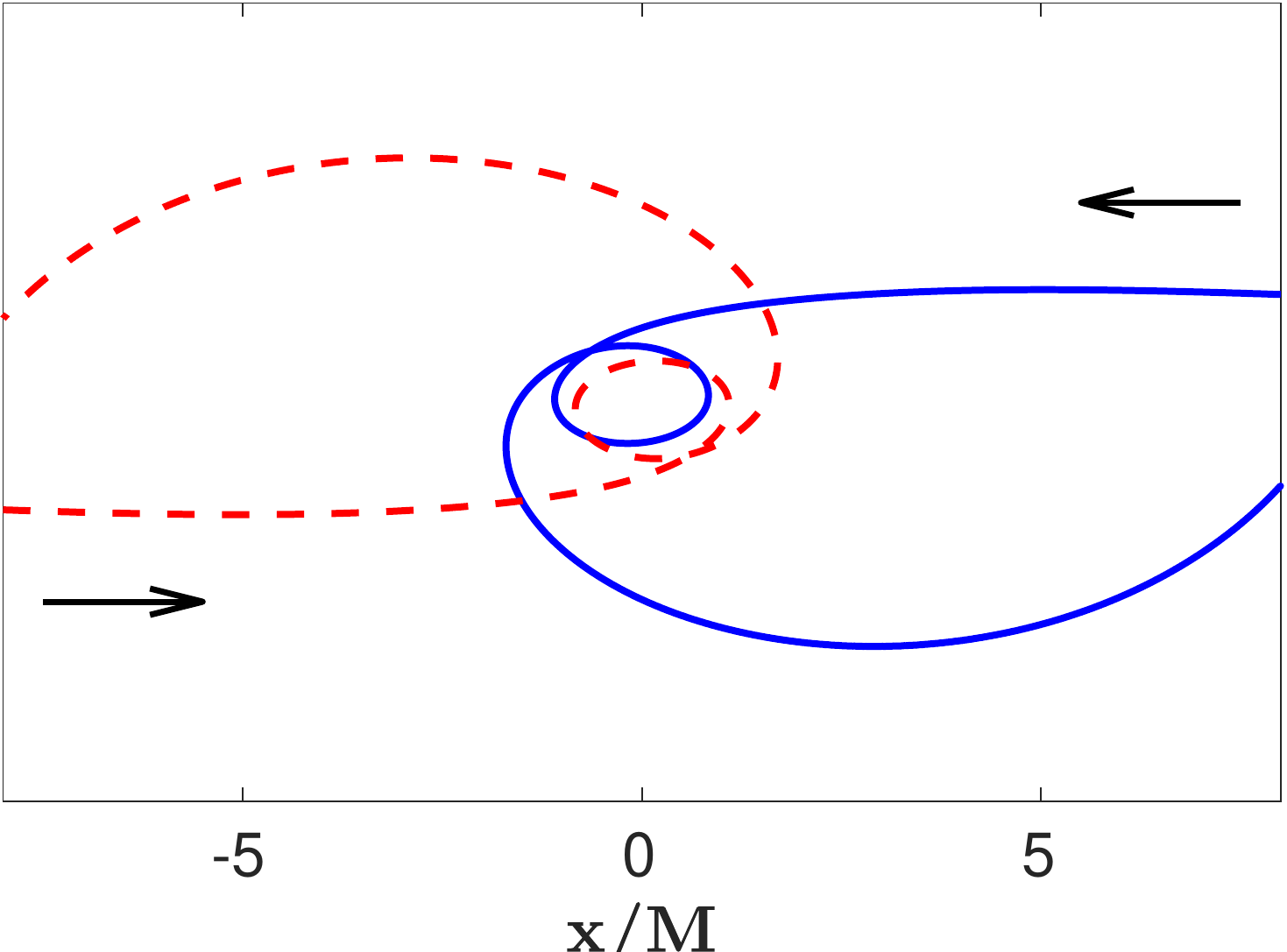}
\hspace{0 cm}
\includegraphics[width=0.333\textwidth,valign=t]{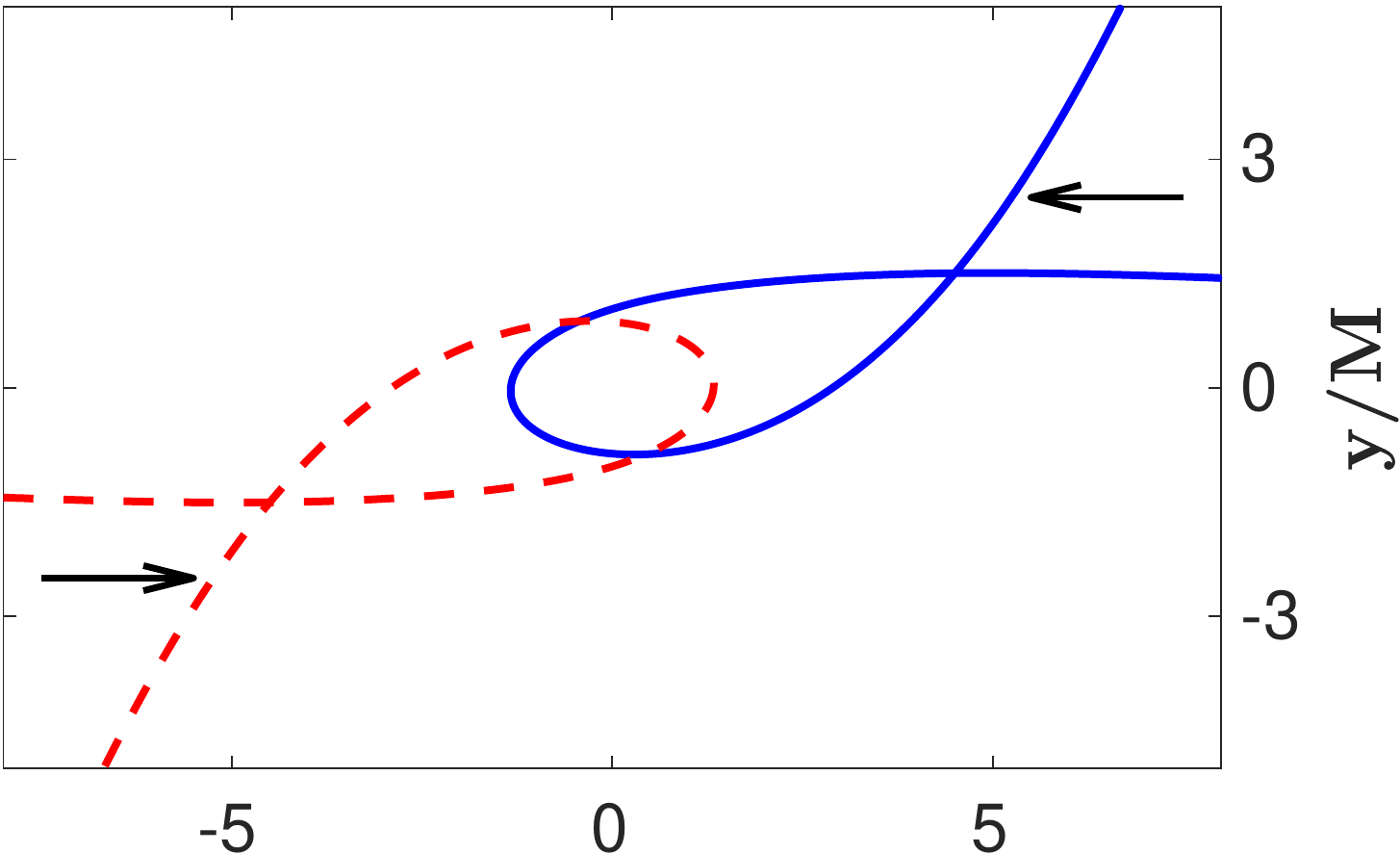}
\caption{Nonscattering (left), marginally scattered (middle), and
  scattered (right) black hole trajectories. Arrows indicate the
  direction each puncture is initially traveling and $M$ is the initial
  ADM mass of the system.} 
\label{fig:trajectories}
\end{figure*}

The chosen set of initial conditions results in a variety of
scattering and nonscattering (i.e. ``merging'') trajectories, as
shown in Fig.~\ref{fig:trajectories}. It is clear from the plots that this
strong-field scattering exhibits much richer morphology than its
classical analogue; some trajectories are
zoom-whirl-like, as can be seen in the left and middle
plots in Fig.~\ref{fig:trajectories}. Likewise, while the right panel does
not show a true ``zoom,'' it is far from Keplerian due to
the emission of gravitational waves and black hole spin-ups.

The plots of Fig.~\ref{fig:trajectories} also relate to how initial
conditions are chosen: we first
selected four Brandt-Br{\"u}gmann momenta 
magnitudes corresponding to initial boosts of $v \approx 0.42, 0.56,
0.66,$ and $0.78$ (as measured in Sec.~\ref{sec:initialboost}). At
each of these magnitudes, we reduced the shooting angle until immediate
merger occurred. This resulted in three classes of trajectories:
immediate mergers (not of interest to this study), marginal scattering
(which may subsequently merge, but the holes remain isolated
long enough for a reliable spin measurement), and scattering; these
are illustrated in the left, middle, and right plots in
Fig.~\ref{fig:trajectories}, respectively. 

\subsection{$v_{\rm boost} = 0.66$ Spin-Up Study}

\begin{figure*}[ht!]
\includegraphics[width=0.49\textwidth]{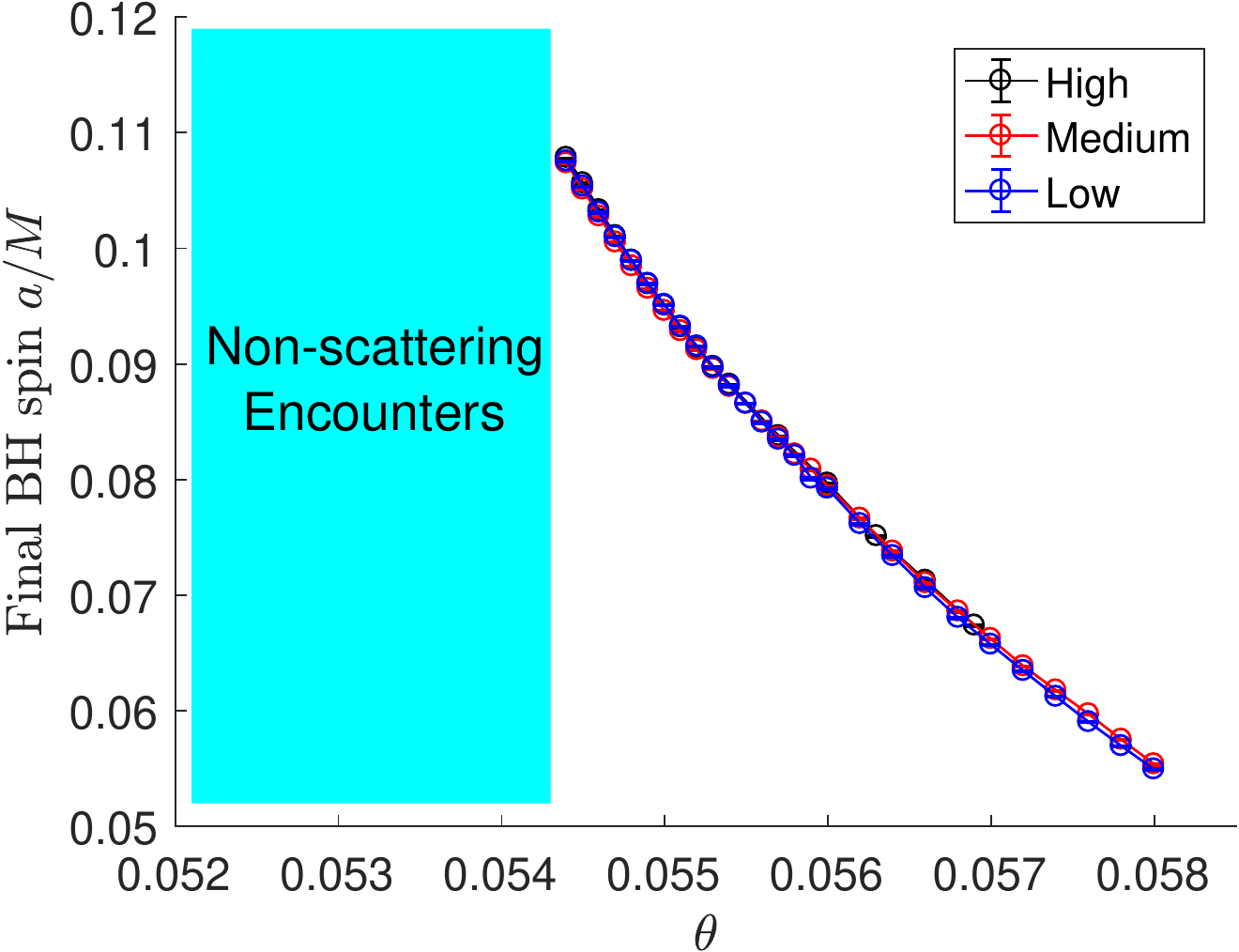}
\includegraphics[width=0.49\textwidth]{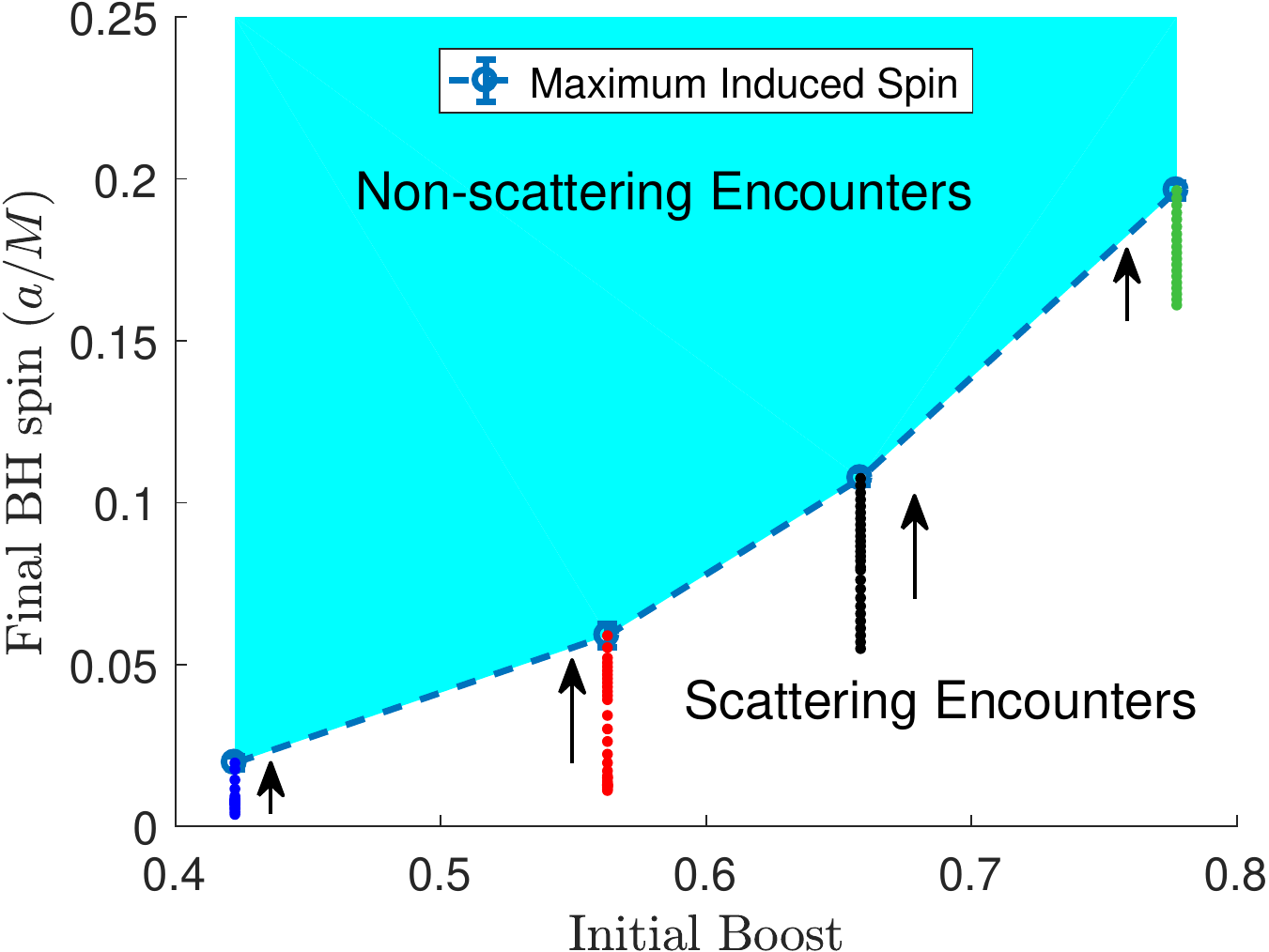}
\caption{{\bf Left panel:} The final spin $a/M$ is computed by
  averaging the Christodoulou spin over late times (once the black
  holes are sufficiently far apart as to be weakly interacting, which
  we have defined as $20M$ separation), and is here plotted against
  the shooting angle $\theta$, which increases
  with impact parameter.\\ 
{\bf Right panel:} The maximum spin-up obtained at a given initial
boost is plotted using open circles; the error bars on these points do not exceed $10\%$ of the point's value. Other spin-ups are displayed as
points; thus, the shooting angle increases downwards on this plot for
a given set of points. The shaded region above the dashed line
indicates the regime of immediate mergers. Black arrows on this plot
indicate the direction of decreasing $\theta$. 
}
\label{spin}
\end{figure*}

The left panel of Fig.~\ref{spin} presents final spin data from the
$v_{\rm boost} = 0.66$ case carried out at three
resolutions. As can be seen, spin-up occurs in \textit{every} case, and
measures at different numerical resolutions agree well with each other: the
relative error between spins at medium and low resolution never
exceeds 1.4\%, and the relative error between high- and medium-resolution
spins never exceeds 0.6\%. Comparing the Christodolou 
and the isolated horizon formalism~\cite{Ashtekar:2004L} measures of
dimensionless spin $J/M^2$ for each hole, we find agreement to within
one percent in the worst case; typically the two measures agree to
three to five significant digits.


The spin-up increases as the shooting angle decreases
towards the separatrix between scattering and nonscattering
trajectories. This increase is nonlinear and concave-up, reaching a maximum
of $0.11$~\footnote{Note that the shooting angle $\theta$ is related to the
impact parameter via $b = d \sin \theta$, where $d$ is the initial
separation between the holes. Since the shooting angles we use
are small, the small-angle approximation gives $b = d \theta$, so
the left panel of Fig.~\ref{spin} would remain almost entirely unchanged
if the impact parameter were plotted instead of the shooting
angle.}.

\subsection{Maximum Spin-Up}

Expanding our analysis to the other boost cases, we again find that in
scattering cases, as the separatrix is approached, the spin-up
increases to its maximum. This study provides the additional insight
that the maximum spin-up depends strongly on the initial boost, as
shown in the right panel of Fig.~\ref{spin}. As we increase the
initial boost of the BHs, the maximum final BH spin 
obtainable at that boost increases; again, this increase is significant
and nonlinear. The maximum spin for $v = 0.42$ was $0.02$, and for
$v=0.56$, we found $0.06$. The maximum induced spin parameter we found
overall was $0.20$, for $v = 0.78$. 

As we demonstrated previously, the spin measures are quite reliable,
and in fact the largest uncertainty in maximum induced spin at a given
initial boost comes from whether or not we truly found the shooting
angle closest to an immediate merger. Therefore, in Fig.~\ref{spin}, we base the error
bars on sampling resolution in shooting angle. (The error bars are
similar in size to the open circles, and become difficult to
distinguish for some of the points.) The error bars in the right panel
of Fig.~\ref{spin} are calculated as the difference between the
highest and second-highest spin measured at a given boost. Again, the
error bars are smaller than the data points themselves.

\subsection{Spin-Up Efficiency}

\begin{figure}[ht!]
\includegraphics[width=0.49\textwidth]{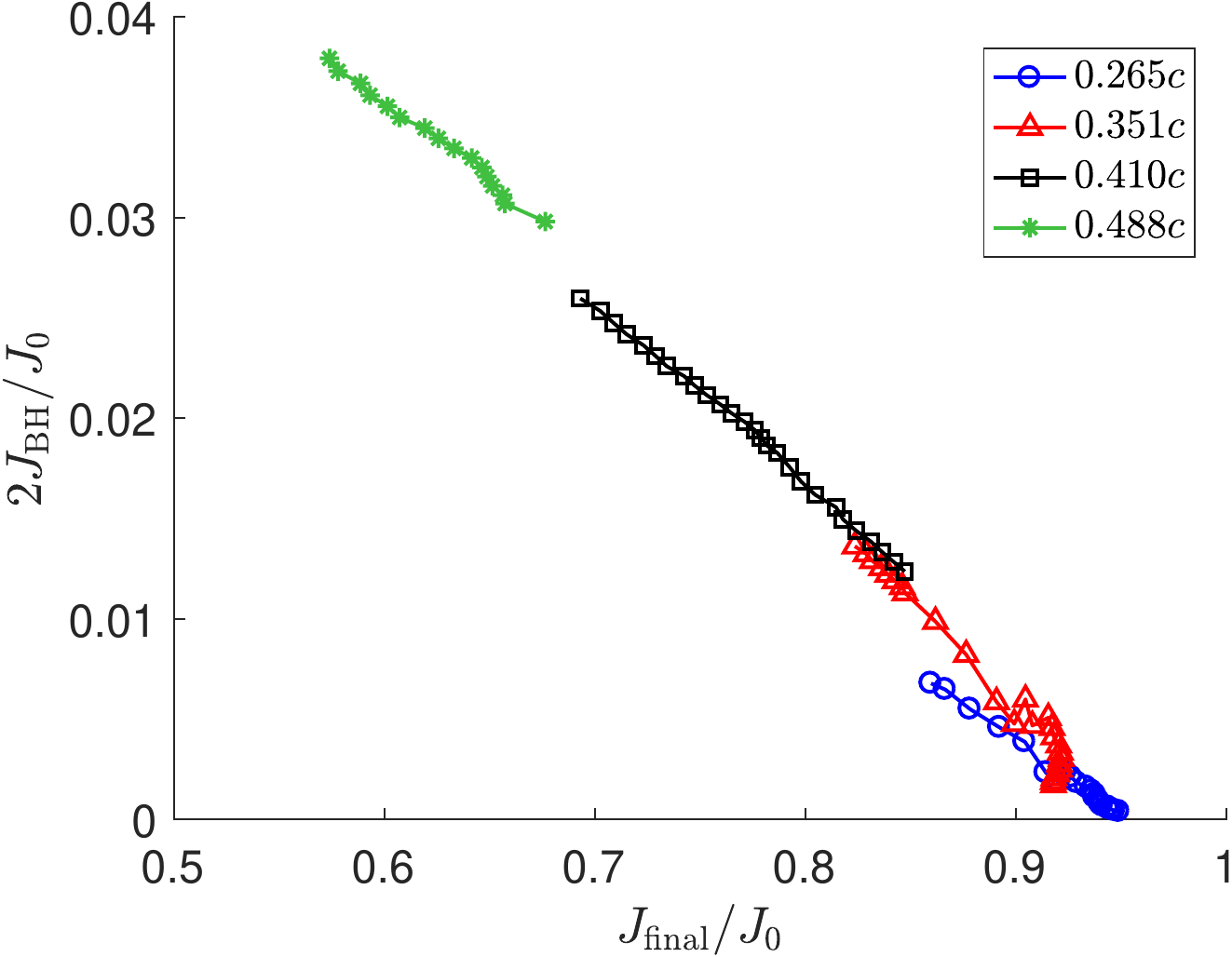}
\caption{Spin-up efficiency, measured as the fraction of the initial
  angular momentum, $J_0$, that was transferred to the spins of the two 
  black holes, $2 J_{\rm BH}$.}
\label{efficiency}
\end{figure}

The angular momentum spinning up the holes originates from
the orbital angular momentum in the initial
data. We would like to identify what fraction of this initial angular momentum is transferred
into spin angular momentum. Figure~\ref{efficiency} plots the
efficiency of spin-up, $2 J_{\rm BH} / J_0$, and the proportion of
orbital angular momentum remaining in the system after the encounter,
$J_{\rm final} = (J_0 - J_{\rm GW} - 2 J_{\rm BH}) /J_0$. Here, we
define $J_{\rm BH}$ to be the dimensionful angular momentum as
calculated using the isolated horizon formalism, and $J_{\rm GW}$ is
defined in Eq.~\ref{eq:JGW}. 

Fixing the initial boost, the smallest angle resulting in a scattering
trajectory corresponds to the largest $J_{\rm BH}$ and, therefore, the largest
spin-up efficiency. Further, we find that higher boosts increase the
spin-up efficiency as well.
We conclude that this spin-up effect has the potential to
significantly impact the orbital evolution of a binary. This is not
taken into account in current, 
commonly used post-Newtonian (PN) approximants, as the spin-up is not
modeled until 4PN order \cite{Poisson:1994}. Further, adding spin-spin
and spin-orbit interaction
terms to initially nonspinning configurations is not yet a standard
approach in PN. Adding such terms would be
necessary to use these results to validate PN theory.

\section{Conclusions}
\label{conc}

We found that decreasing the shooting angle at a given initial boost
increases the spin induced onto black holes that undergo a scattering
trajectory. Further, as the initial boost is increased,
the maximum spin-up and spin-up efficiency (i.e. the fraction of the
initial angular momentum transferred to the black holes' spins)
increases nonlinearly. 

The maximum spin-up observed was $a/M=0.20$ imparted on each hole,
which occurs with an initial boost of $0.78c$ (the maximum
boost chosen). This also corresponds to the maximum spin-up efficiency
observed, of $2J_{BH}/J_0 = 3.9\%$. Once
post-Newtonian theory has been completed at 4PN order for this type of interaction, this work will provide an
exciting new avenue for validating PN theory directly with numerical
relativity calculations. 

These results indicate that the spin-ups and spin-up efficiencies may
increase significantly with larger initial boosts, which will be a focus of
future work. However, this work presents boosts near the upper limit
allowed by conformally flat Brandt-Br{\"u}gmann initial data, so
larger initial boosts will require improved initial data {\it a
  la} Ruchlin, Healy, Lousto, and
Zlochower~\cite{Ruchlin:2014}, as well as higher
numerical resolutions.

This is a very rich problem with a massive parameter space left to
explore, and we hope to eventually study cases with varying initial
spins and unequal mass ratios. With so many options available, we plan
to add black hole scattering experiments to
BlackHoles@Home, a distributed computing project enabling
numerical relativity calculations of black hole interactions to be
performed on consumer-grade
desktop or laptop computers~\cite{Ruchlin:2017,NRPy,SENR}.

In future work we would also like to extract more 
gauge-invariant quantities from the gravitational-wave data, such as
the peak frequency of the radiation, $f_{\rm peak}$. We attempted to
do so with these data, but they were too noisy to obtain reliable
measures. It would also be useful as an additional validation to
measure the conservation of total $J$ in the system by, e.g., directly
computing the ADM $J_{\rm final}$ and comparing it to the proxy
$J_0-J_{\rm GW}-2J_{\rm BH}$ that we used. Verifying conservation of total
$M_{\rm ADM}$ will be useful as well.

More broadly speaking, we hope that the calculations performed
here will allow us to characterize a subtle strong-field
effect that could nonetheless bring about observable effects in future
high-precision gravitational-wave observations.

\section*{Acknowledgments}
We thank I.~Ruchlin for helpful discussions. This work was supported
by NSF awards OIA-1458952, PHY-1607405 and PHY-1912497, as well as
NASA awards ISFM-80NSSC18K0538 and TCAN-80NSSC18K1488. Computational
resources were provided by West Virginia University's Spruce Knob
high-performance computing cluster, funded in part by NSF EPSCoR
Research Infrastructure Improvement Cooperative Agreement No. 1003907,
the state of West Virginia (WVEPSCoR via the Higher Education Policy
Commission), and West Virginia University.

\bibliographystyle{apsrev4-1}
\bibliography{inducedSpinPaper}

\end{document}